\documentclass[aps,prl,twocolumn,floatfix,noshowpacs,tightenlines,noshowkeys,superscriptaddress,amsmath,amssymb,nofootinbib]{revtex4}
\usepackage{amssymb,amsbsy,epsfig,color,graphicx}
\usepackage{color}
\usepackage{[longtable}
\usepackage{array}
\usepackage{dcolumn}   
\usepackage{cellspace}
\usepackage{mathtools}
\usepackage{amstext}
\usepackage{amssymb}
\usepackage{stmaryrd}
\usepackage{stackrel}
\usepackage{graphicx}
\usepackage{esint}
\usepackage[utf8]{inputenc}
\usepackage{blindtext}
\usepackage{float}
\restylefloat{table}
\usepackage{booktabs}
\usepackage{enumitem}

\usepackage{etoolbox} 
\usepackage{lipsum} 
\usepackage[capitalize]{cleveref}
\usepackage{multirow}
\usepackage[caption=false]{subfig}
\renewcommand\[{\begin{equation}}
\renewcommand\]{\end{equation}}

\newcommand{\ba}{\begin{eqnarray}}
\newcommand{\ea}{\end{eqnarray}}

\makeatletter

\appto{\appendix}{%
\@ifstar{\def\theequation@prefix{A.}}%
{}%
}
\makeatother

\begin{document}

\title{Nonlocality amplifies echoes}

\author{Luca Buoninfante}
\affiliation{Van Swinderen Institute, University of Groningen, 9747 AG, Groningen, The Netherlands}
\affiliation{Dipartimento di Fisica "E.R. Caianiello", Universit\`a di Salerno, I-84084 Fisciano (SA), Italy}
\affiliation{INFN - Sezione di Napoli, Gruppo collegato di Salerno, I-84084 Fisciano (SA), Italy}

\author{Anupam Mazumdar}
\affiliation{Van Swinderen Institute, University of Groningen, 9747 AG, Groningen, The Netherlands}

\author{Jun Peng}
\affiliation{Van Swinderen Institute, University of Groningen, 9747 AG, Groningen, The Netherlands}

\begin{abstract}
In this paper we will provide smoking-gun signatures of nonlocal interactions while studying reflection and transmission of waves bouncing through two Dirac delta potentials. In particular, we will show that the transmission of waves is less damped compared to the local case, due to the fact that nonlocality weakens the interaction. As a consequence the echoes are amplified. These signatures can be potentially detectable in the context of gravitational waves, where two Dirac delta potentials can mimic the two potential barriers at the surface and at the photon sphere of an ultra compact object, or, at the two photon spheres of a wormhole, experiencing nonlocal interactions.
\end{abstract}

\maketitle


The nonlocal interactions have been widely studied since the days of Yukawa~\cite{Yukawa:1950eq}, Pias and Uhlenbeck~\cite{Pais:1950za}. It has been studied widely in the context of quantum field theory in order to ameliorate the ultraviolet (UV) (or, in other words, short-distance) behavior of loop integrals and scattering amplitudes~\cite{Efimov:1967pjn,Tomboulis:1997gg,Moffat:1990jj,Buoninfante:2018mre}. One of the striking features of nonlocality is that it weakens the interaction, and as a direct consequence it can regularize singularities through smearing out a point-like Dirac delta distribution~\cite{Siegel:2003vt,Tseytlin:1995uq,Biswas:2005qr,Biswas:2011ar}.  The simplest nonlocal interaction can be captured by infinite derivative field theories, where the kinetic operator is generalized by means exponentials of {\it entire} functions, which do not introduce any new dynamical degrees of freedom, and modifies the UV behavior of the theory. The latter criteria guarantees perturbative unitarity and classical stability which, in general, are absent in any finite higher derivative theory, see~\cite{Tomboulis:1997gg,Biswas:2005qr,Biswas:2011ar,Woodard:2015zca}.  All these interesting results have inspired the generalization of the Einstein-Hilbert action for gravity to infinite derivative gravity (IDG), with form factors containing nonlocal operators in order to make the theory well defined both at classical and quantum level~\cite{Tomboulis:1997gg,Tseytlin:1995uq,Siegel:2003vt,Biswas:2005qr,Moffat:2010bh,Biswas:2011ar}. In a recent generalization which contains all possible curvature corrections~\cite{Biswas:2011ar}, it was shown that IDG can resolve the point-like singularity in a static~\cite{Biswas:2011ar,Buoninfante:2018rlq,Buoninfante:2018xiw,Edholm:2016hbt,Boos:2018bxf}, and the ring singularity in a rotating spacetime metric~\cite{Buoninfante:2018xif}, and with the presence of torsion as well~\cite{delaCruz-Dombriz:2018aal}. In a dynamical collapse also there are no formation of horizon and singularities~\cite{Frolov:2015bia}. Furthermore, such gravitational actions also help in order to understand the cosmological Big Bang singularity problem~\cite{Biswas:2005qr,Biswas:2010zk}, and potentially resolving the event horizon in the context of astrophysical compact objects~\cite{Buoninfante:2019swn}.

The aim of this paper is to provide a smoking gun signature for nonlocal interactions. In particular, we will study the propagation of waves in nonlocal massless scalar field theory in presence of two Dirac delta potential barriers in one spatial dimension. We will work out the quasi-normal modes (QNMs) and the echoes of an initial incoming pulse and make the comparison with the local case. Our analysis allows to make a very powerful analogy in astrophysics. First of all, the two Dirac deltas can be used to model the two symmetric photon spheres of a wormhole \cite{Bueno:2017hyj}. In the case of a static spacetime metric, the double delta potential could also mimic the two potential barriers at the surface and at the photon sphere of an ultra compact object (UCO), located at $2GM(1+\epsilon)$ and $3GM,$ respectively, with $\epsilon < 0.5,$ where $G$ is Newton's constant and $M$ the mass of the UCO~\cite{Cardoso:2016oxy}. Scenarios like this have also appeared in the literature in many string inspired realizations in order to avoid the information-loss paradox~\cite{Mathur:2005zp,Giddings:2006sj,Unruh:2017uaw,Buoninfante:2019swn}. Hence, despite the simplicity of the model, we will be able to capture the main features of nonlocality which turn out to be universal and we believe will also hold in a more realistic astrophysical scenarios.

Let us consider a nonlocal massless scalar field $\Phi$ interacting with a potential $V$ in $1+1$ dimensions with a nonlocal operator
given by an exponential of an {\it entire} function
\begin{equation}
[F(\Box)\Box-V]\Phi=0\,,~~~F(\Box)=e^{-\ell_s^2\Box}\,;
\label{scalar-field-eq}
\end{equation}
we will work with the simplest kinetic operator above but other choices are also possible \cite{Tomboulis:1997gg,Modesto:2011kw,Edholm:2016hbt,Boos:2018bxf}.
In Eq.\eqref{scalar-field-eq}, $\ell_s$ is the fundamental length scale of nonlocality below which new physics should manifest and $\Box=-\partial_t^2+\partial_x^2$ is the flat d'Alembertian operator. Note that we adopt the metric signature; $\eta={\rm diag}(-1,1,1,1)$, and use the Natural Units $\hbar=1=c\,.$ By separating the time and spatial variables with the ansatz $\Phi(t,x)=e^{-i\omega t}\psi(x),$ Eq.\eqref{scalar-field-eq} becomes \cite{Boos:2018kir}
\begin{equation}
\left[e^{-\ell_s^2(\partial_x^2+\omega^2)}(\partial_x^2+\omega^2)-V(x)\right]\psi(x)=0\,,\label{schrodinger}
\end{equation}
which is a nonlocal version of the Schr\"odinger equation. For general potential barriers, $V(x)$ is assumed to be positive and to satisfy $V(x)\to0$ as $x\to\pm\infty,$ so that the general solution of $\psi(x)$ has the form
\begin{equation}
\psi(x)=\begin{cases}
	Ae^{i\omega x}+Be^{-i\omega x},\quad x\to-\infty,\\
	Ce^{i\omega x},\quad x\to+\infty.
\end{cases}
\end{equation}
The reflection and transmission coefficients, $\mathcal{R}$ and $\mathcal{T}$ respectively, are defined as
\begin{equation}
\mathcal{R}={B}/{A}\,,\quad \mathcal{T}={C}/{A}\,,\quad |\mathcal{R}|^2+|\mathcal{T}|^2=1\,.
\end{equation}
For a general potential Eq.\eqref{schrodinger} may not be easily solvable, but there exist analytic solution in some cases. Here, we assume a symmetric double Dirac delta potentials
\begin{equation}
V(x)=\lambda \delta(x+a)+\lambda \delta(x-a)\,.\label{double-delta}
\end{equation}
Note that a single Dirac delta potential in nonlocal theory was first studied in \cite{Boos:2018kir}. The nonlocal Schr\"odinger equation \eqref{schrodinger} with the potential \eqref{double-delta} can be solved analytically by using the Lippmann-Schwinger method \cite{Lippmann:1950zz}
\begin{equation}
\psi(x)=\psi_0(x)-\int^\infty_{-\infty} dx'G(x,x')V(x')\psi(x')\label{LSequation0}
\end{equation}
where $G(x,x')$ is the Green function defined through the equation $e^{-\ell_s^2(\partial_x^2+\omega^2)}(\partial_x^2+\omega^2)G(x,x')=-\delta(x-x'),$ and reads \cite{Boos:2018kir}
\begin{equation}
\begin{array}{rl}
{\cal G}(z)=&\displaystyle \frac{i}{4\omega}[e^{i\omega z}{\rm Y}(z)+e^{-i\omega z}{\rm Y}(-z)],\\
{\rm with}&\displaystyle {\rm Y}(z)=1+{\rm erf}\left(i\omega \ell_s+\frac{z}{2\ell_s}\right),
\end{array}
\end{equation}
which in the limit $\ell_s\rightarrow 0$ reduces to the local Green function ${\cal G}_0(z)=\frac{i}{2\omega}e^{i\omega|z|}\,,$ as expected. The Green function turns out to be symmetric, $G(x,x')=G(x',x)$, and depends only on the variable $z=x-x':$ $G(x,x')\equiv {\cal G}(z)$, ${\cal G}(z)={\cal G}(-z)\,.$ Hence, by using the Lippmann-Schwinger equation \eqref{LSequation0}, the solution for the symmetric double delta in \eqref{double-delta} can be shown to be
\begin{equation}
\psi(x)=e^{i\omega x}-\lambda\sum_{j,k=1}^2{\cal G}(x-a_k)\Phi^{-1}_{kj}e^{i\omega a_j},\label{solutionpsi}
\end{equation}
where
\begin{equation}
\Phi_{kj}=\begin{cases}
	\displaystyle 1+\frac{i\lambda_j }{2\omega}[1+{\rm erf}(i \omega \ell_s)],\quad k=j;\\
	\displaystyle \frac{i\lambda_j }{4 \omega}[e^{i 2\omega a}{\rm Y}(2a)\!+\!e^{-i2 \omega a}{\rm Y}(-2a)],\quad  k\ne j\,,
\end{cases}\label{solution}
\end{equation}
and we have defined $a_1=a,$ $a_2=-a\,.$ By assuming that the incoming wave comes from the left, we can find the reflection and transmission coefficients for the solution in Eq.\eqref{solution} by identifying the coefficients in front of $e^{-i\omega x}$ when $x\ll a$ and of $e^{i\omega x}$ when $x\gg a\,,$ respectively; they are given by
\begin{widetext}
\begin{eqnarray}
\mathcal{R}=&&-\frac{2[-2\frac{\omega}{\lambda} i+1+{\rm erf}(i \omega \ell_s)]{{\rm cos}(2 \omega a)}-[1+ {\rm erf}(i \omega \ell_s+\frac{a}{\ell_s})]e^{2 i \omega a}-[1+{\rm erf}(i \omega \ell_s-\frac{a}{\ell_s})]e^{-2 i \omega a}}{[-2\frac{\omega}{\lambda} i+1+{\rm erf}(i \omega \ell_s)]^2-\frac{1}{4}\{[1+ {\rm erf}(i \omega \ell_s+\frac{a}{\ell_s})]e^{2 i \omega a}+[1+{\rm erf}(i \omega \ell_s-\frac{a}{\ell_s})]e^{-2 i \omega a}\}^2},\label{reflec}\\[2mm]
\mathcal{T}=&&1-
\frac{2[-2\frac{\omega}{\lambda} i+1+{\rm erf}(i \omega \ell_s)]-\{[1+{\rm erf}(i \omega \ell_s+\frac{a}{\ell_s})]e^{2i\omega a}+[1+{\rm erf}(i \omega \ell_s-\frac{a}{\ell_s})]e^{-2i\omega a}\}\cos(2\omega a)}{[-2\frac{\omega}{\lambda} i+1+{\rm erf}(i \omega \ell_s)]^2-\frac{1}{4}\{[1+ {\rm erf}(i \omega \ell_s+\frac{a}{\ell_s})]e^{2 i \omega a}+[1+{\rm erf}(i \omega \ell_s-\frac{a}{\ell_s})]e^{-2 i \omega a}\}^2},\label{transm}
\end{eqnarray}
\end{widetext}
and one can easily check that $|\mathcal{R}|^2+|\mathcal{T}|^2=1$.

\textbf{\emph{QNMs}:} The poles of $\mathcal{R}$ and $\mathcal{T}$ describe the QNMs of the system, which satisfy the boundary condition such that there is no incoming wave at infinity, i.e. $\psi\to e^{-i\omega x}$ as $x\to-\infty$ and $\psi\to e^{i\omega x}$ as $x\to+\infty$ \cite{Boonserm:2010px}. In a realistic astrophysical framework, the quantities in \eqref{reflec} and \eqref{transm} are crucial for understanding the effect of nonlocality in the UCOs.  We can check that the poles of the reflection and transmission coefficients are given by the equations
\begin{eqnarray}
2[-2\frac{\omega}{\lambda} i+1+{\rm erf}(i \omega \ell_s)]=&\pm\left\lbrace [1+ {\rm erf}(i \omega \ell_s+\frac{a}{\ell_s})]e^{2 i \omega a}\right.\nonumber\\
&\!\!\!\!\left.+[1+{\rm erf}(i \omega \ell_s-\frac{a}{\ell_s})]e^{-2 i \omega a}\right\rbrace, \label{qnf}\nonumber\\
\end{eqnarray}
whose roots are of the type $\omega_n=\mathcal{W}_n-i\omega_n^{\prime}$ and are called quasi-normal frequencies (QNFs). Note that for $\omega$ with a negative imaginary part, the time dependent piece $e^{-i\omega t}$ behaves like a damped oscillation: the real part gives the energy carried by the wave and the imaginary part (inverse) gives the time scale over which the wave decays after interacting with the potential barrier. It is worthwhile mentioning that in the limit, $a=0$, we recover the quasi-normal condition for a single Dirac delta potential with strength $2\lambda,$ i.e. $1-i\omega/\lambda+{\rm erf}(i\omega \ell_s)=0$ \cite{Boos:2018kir}; while in the limit $\ell_s=0\,,$ Eq.\eqref{qnf}  reduces to the well known local case, $1-i2\omega/\lambda=\pm e^{2i\omega a}\,.$
\begin{figure}[t!]
	\includegraphics[scale=0.395]{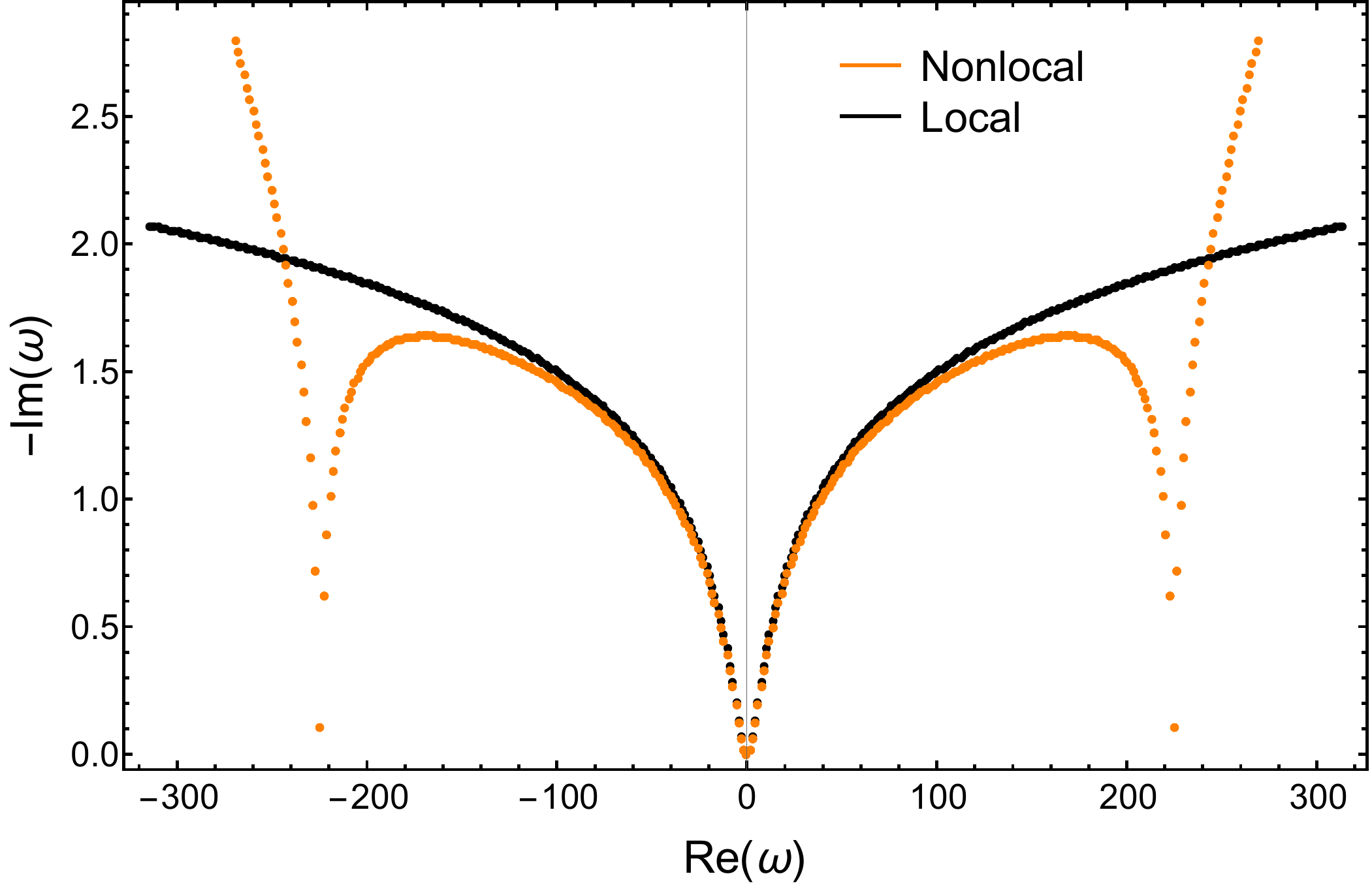}\quad\includegraphics[scale=0.395]{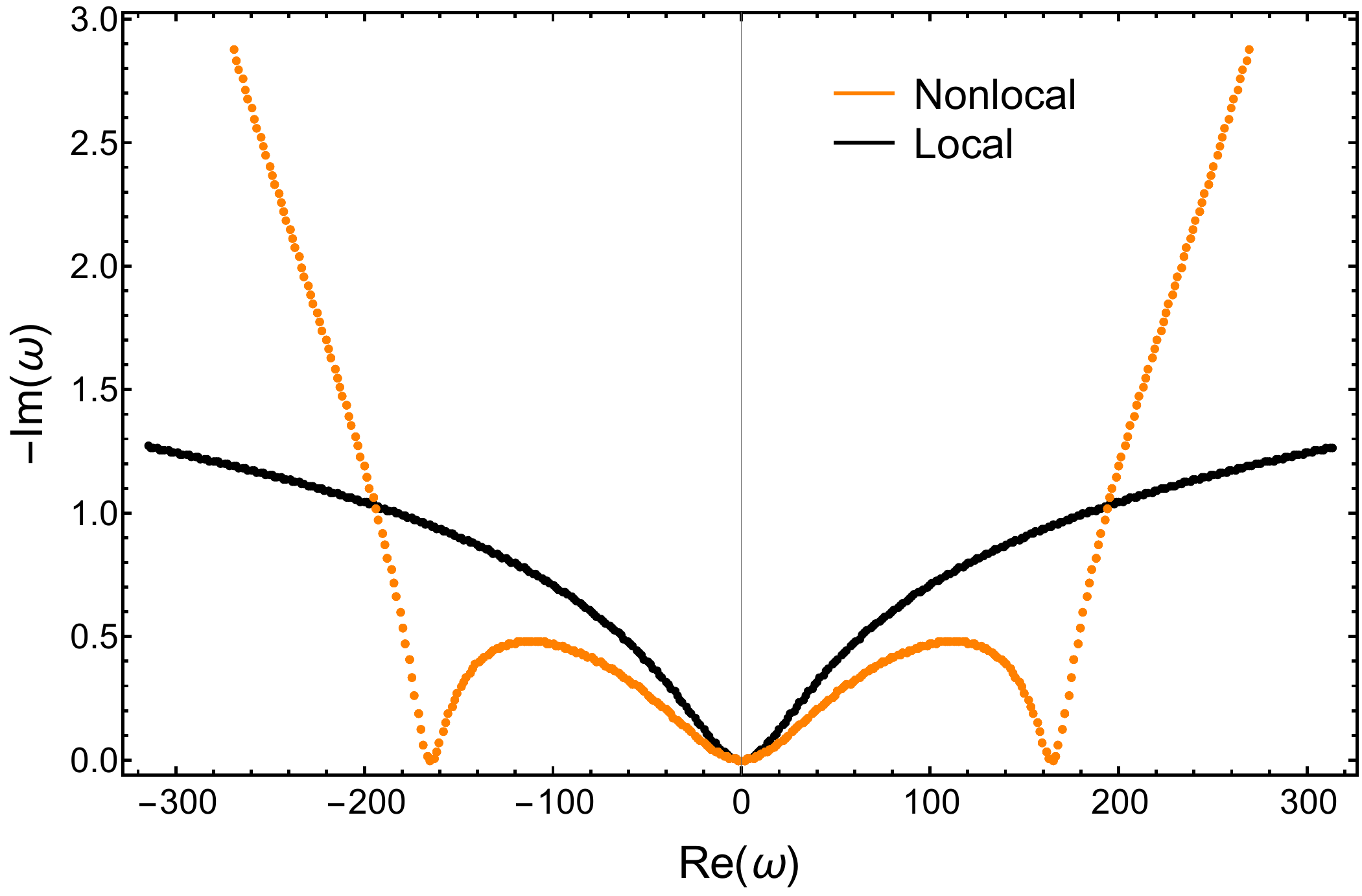}
	\caption{We have shown the behaviour of the real part of the QNFs $\mathcal{W}_n$ as a function of the negative imaginary part $\omega_n^{\prime}$ for both local (black line) and nonlocal (orange line) cases; the latter was obtained by solving numerically Eq.\eqref{qnf2}. In the upper panel we have set $a=1,$ $\ell_s=0.01$ and $\lambda=10,$ while in the bottom panel $a=1,$ $\ell_s=0.01$ and $\lambda=50\,.$}
	\label{fp-plot1}
\end{figure}

Generally, Eq.\eqref{qnf} cannot be solved, but we can simplify it and obtain an analytic solution by working in a specific regime which turns out to be the most interesting one from a physical point of view. Since we expect the fundamental scale of nonlocality $\ell_s$ to be always smaller than the distance between the surface and the photon sphere of a UCO, see~\cite{Cardoso:2016oxy,Buoninfante:2019swn}, it is sensible to require $a\gg\ell_s\,.$ Moreover, when studying the QNMs and the echoes produced by an initial pulse (see also below) the low frequencies are the most relevant ones as they correspond to longer damping times, therefore we can also impose the inequality $|\omega| \ell_s<a/\ell_s\,.$ In such limits, ${\rm erf}(i\omega \ell_s +a/\ell_s)\rightarrow 1$ and ${\rm erf}(i\omega \ell_s -a/\ell_s)\rightarrow -1\,,$ so that Eq.\eqref{qnf} reduces to
\begin{equation}
1-i2\frac{\omega}{\lambda} +{\rm erf}(i\omega \ell_s)=\pm e^{2i\omega a},\label{qnf2}
\end{equation}
which in the limit $|\omega|\ell_s\ll 1$ can be further simplified since ${\rm erf}(i\omega \ell_s)\approx \frac{2}{\sqrt\pi} i\omega \ell_s\,.$ Indeed, in this regime we can find an analytic expression for the QNMs in terms of the Lambert $\rm{W}$ function:
\begin{equation}
\omega_n=i\left[\frac{{\rm W}_n(\pm\alpha ae^{\alpha a})}{2a}-\frac{\alpha}{2}\right],\quad \frac{1}{\alpha}=\frac{1}{\lambda}-\frac{\ell_s}{\sqrt\pi} .\label{analytic-sol}
\end{equation}
Furthermore, we can solve numerically Eq.\eqref{qnf2} and study the behaviour of the real part of the QNMs, ${\rm Re}(\omega)=\mathcal{W}_n\,,$ as a function of the negative imaginary part $-{\rm Im}(\omega)=\omega_n^{\prime}\,.$ From Fig. $1$ we can see that nonlocal effects make the imaginary part of the QNFs smaller, i.e. the damping time is longer. The closer the ring-down energy $\mathcal{W}_n\,$ is to the nonlocal energy scale $1/\ell_s\equiv M_s$, the more relevant nonlocality is, which means the more obvious the difference of imaginary parts between local and nonlocal is.  Especially, when the strength of the delta potential $\lambda\gtrsim 1/\ell_s\equiv M_s$ , the damping is smaller. Therefore, a crucial implication due to nonlocality is that low frequency waves survive a {\it longer} life-time as compared to the local case. This feature was expected since it is known that nonlocality weakens the interaction \cite{okada,Buoninfante:2018mre}. Moreover, when ${\cal W}_n\ell_s\sim{\cal O}(1)$, we can also notice the presence of a turning point after which the damping time decreases. This property is mathematically related to the structure of the error function with complex argument but it is not relevant from a physical point of view since it will only describe short-lived waves with energies ${\cal W}_n \ell_s\gg1\,.$

\textbf{\emph{Echoes}:} So far we have learned that nonlocal effects weaken the interaction between wave and potential barriers, and as a consequence the time scales over which a transmitted wave decays turn out to be longer than the local case. We now wish to use the spectral features of the QNMs and study the consequences induced by nonlocality on the echoes production. In order to do so we need to work in the time domain where the form of the wave can be cast as an infinite linear combination of QNMs \cite{Bueno:2017hyj,Mark:2017dnq}:
\begin{equation}
\Psi(t)=\sum^{\infty}_{n=-\infty}c_ne^{-i\omega_nt}\,,\label{signal}
\end{equation}
which in a more realistic scenario can represent the signal produced at the photon sphere after relaxation. Because the oscillation period of the real parts of the QNMs is ${\rm Re}(\omega_{n+1}-\omega_n)\simeq \pi/2a\,,$ which is the same also in the local case, and since ${\rm Im}(\omega_n)\ll {\rm Re}(\omega_n),$ the function $\Psi(t)$ is in good approximation periodic, with period $T=4a\,.$ This means that the primary signal is reproduced periodically, as it can be reflected by the second potential barrier (UCO's surface), thus giving rise to echoes.

This feature allows us to express the coefficients $c_n$ as a Fourier transform of the first echo, $\Psi^{(0)}_{1^{\rm st}{\rm echo}}(t),$
\begin{equation}
c_n=\frac{1}{4a}\int^{4a}_0dt\Psi^{(0)}_{1^{\rm st}{\rm echo}}(t)e^{i\omega_nt}\,.\label{c_n}
\end{equation}
We can choose the first echo as a Gaussian wave packet of the form
\begin{equation}
\Psi^{(0)}_{1^{\rm st}{\rm echo}}(t)=e^{-i\omega_0(t-t_0)}e^{{-(t-t_0)^2}/{2\tau^2}},
\end{equation}
where $\omega_0$ is the leading frequency of the primary signal. Assuming that $\tau\ll 2a$ we can obtain an approximation for the coefficients $c_n$ of the subsequent echoes by computing the integral in Eq.\eqref{c_n} \cite{Bueno:2017hyj}
\begin{equation}
c_n\simeq \sqrt{\frac{\pi}{2}}\frac{\tau}{2a}\exp\left[i\omega_nt_0-\frac{\tau^2}{2}(\omega_n-\omega_0)^2\right]\,.\label{coefficients}
\end{equation}
\begin{figure}[t!]
	\includegraphics[scale=0.243]{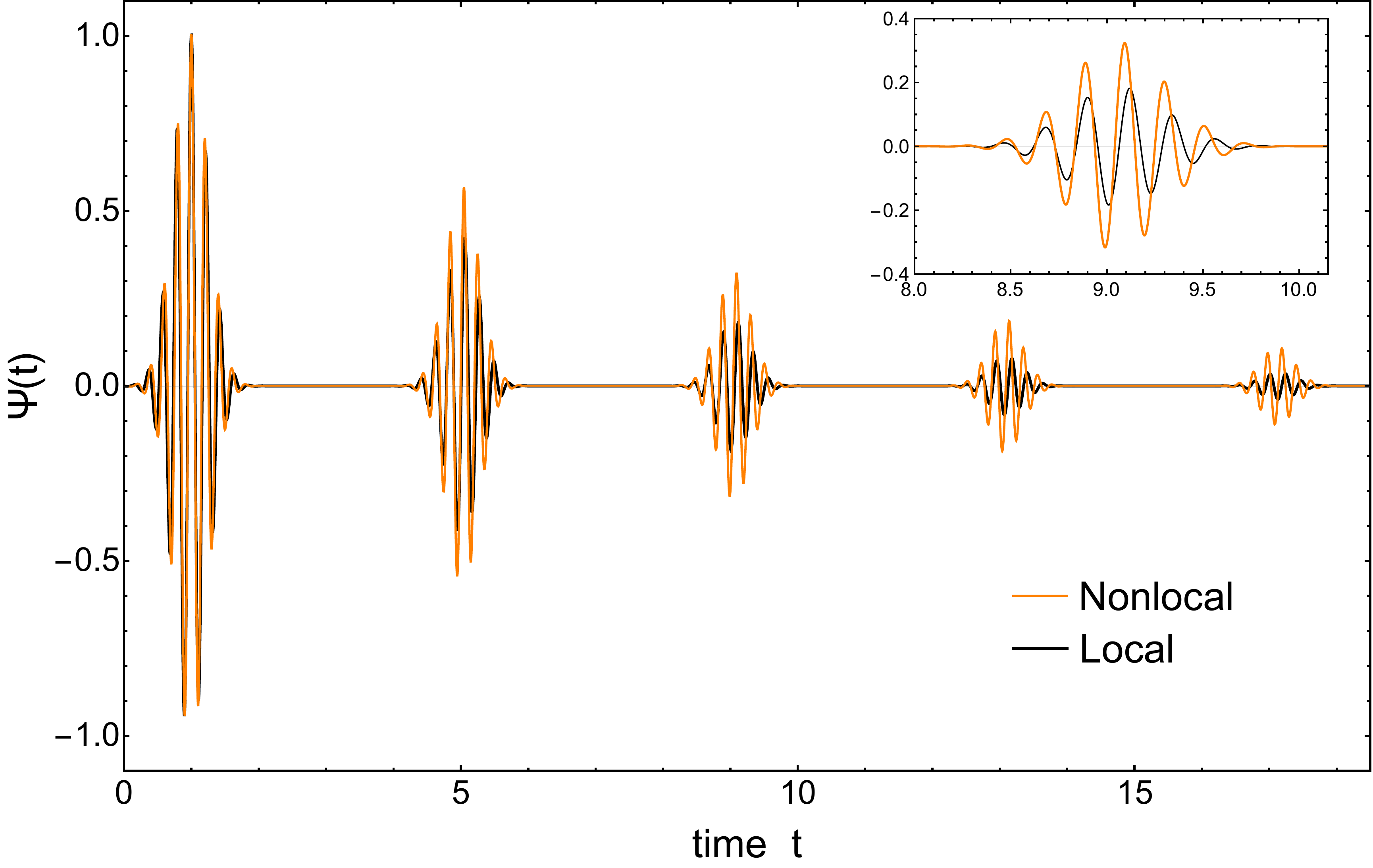}
	\caption{We have plotted the behaviour of the signal in Eq.\eqref{signal}, with coefficients \eqref{coefficients}, which describes the evolution of the echoes produced at a fixed spatial point, for both local (black line) and nonlocal (orange line) cases. We have chosen $t_0=1,$ $\tau=0.25,$ and set $a=1,$ $\lambda=50,$ $\ell_s=0.01\,.$ }
\end{figure}
Hence, we can now use Eq.\eqref{signal} with the coefficients given in Eq.\eqref{coefficients} to plot the time dependent profile of the signal, see Fig. $2.$ Note that, because only the long-lived QNMs (QNFs with small negative imaginary part) led to the echoes in the late time \cite{Oshita:2018fqu}, thus it is also sufficient to analyze these results using Eq.\eqref{analytic-sol}.

Before we conclude, let us capture the salient features of our analysis. In the regime $\lambda\ll 1/\ell_s\equiv M_s$, we have already seen that the QNMs are very similar to the local case, therefore the same applies for the echoes. It means that when the height of the potential barriers is below the energy scale $1/\ell_s\equiv M_s$, nonlocality does not play a relevant role, i.e. the echoes are almost the same as the local case. However, when the height of the potential barrier is either comparable or larger than the nonlocal energy scale, i.e. $\lambda \gtrsim M_s,$ the damping time of the nonlocal case is much longer at low frequency. So a very distinct difference arises between local and nonlocal cases. As shown in Fig. $2\,,$ the amplitudes of the late time echoes are amplified with respect to the local ones, meaning that nonlocality acts as an {\it amplifier}. Actually, we can adjust the degree of amplification by adjusting the height of the potential $\lambda$. Such a behavior is consistent with the fact that the nonlocal interaction makes the life-times of the waves longer~\cite{Buoninfante:2019swn}, which means they are less damped in time, therefore the echo signal is more amplified. The presence of a smaller damping scale also implies that in the process of transmission less energy is dissipated when the interaction between the wave and the potential barrier is nonlocal, therefore the echoes are louder.

Let us conclude the paper by emphasizing that nonlocal effects can be amplified in presence of double delta potential barriers, which can be experimentally probed. In our opinion, gravitational waves in the merger of two UCOs, or wormholes, might yield the secret of nonlocality, which goes in conjunction with the avoidance of an event horizon to explain that blackhole information loss paradox. It is worthwhile mentioning that this kind of nonlocal effects could be captured in models of condensate matter system, in which the reflection and transmission properties of quantum mechanical barriers can be potentially tested in a table-top experiment. For a different application, see also Ref.\cite{Belenchia:2015ake} were nonlocal effects on quantum mechanical oscillators were studied.

Our analysis has been made in the simple case of flat background but future studies are needed to take into account the presence of a curvature and of a more realistic effective potential, but we believe that the essence will be very similar to what we have presented here. The key smoking-gun signature will be the presence of a smaller damping scale and the amplification of echoes, which is a distinctive feature of nonlocality which is surely absent in any local theory.

\paragraph{Acknowledgements.} AM's research is financially supported by Netherlands Organization for Scientific Research (NWO) grant number 680-91-119. JP is financially supported by China Scholarship Council.

\end{document}